\documentclass[reprint,aps, prb, amsmath, amssymb, longbibliography,superscriptaddress]{revtex4-2}
\usepackage{bm}
\usepackage{graphicx}
\def\be{\begin{equation}}
\def\ee{\end{equation}}
\def \bea{\begin{eqnarray}}
\def \eea{\end{eqnarray}}
\def \nn{\nonumber}
\DeclareUnicodeCharacter{2212}{\textminus}
\usepackage[colorlinks, linkcolor= blue, citecolor = blue, urlcolor=blue]{hyperref}
\newcommand{\vect}[1]{\boldsymbol{#1}}

\begin{document}

\title{Band geometric transverse current driven by inhomogeneous AC electric field} 
\author{M. Maneesh Kumar}
\thanks{Maneesh and Sanjay contributed equally to this manuscript.}
\affiliation{Department of Physics, Indian Institute of Technology Kanpur, Kanpur 208016}
\author{Sanjay Sarkar}
\thanks{Maneesh and Sanjay contributed equally to this manuscript.}

\affiliation{Department of Physics, Indian Institute of Technology Kanpur, Kanpur 208016}

\author{Amit Agarwal}
\email{amitag@iitk.ac.in}
\affiliation{Department of Physics, Indian Institute of Technology Kanpur, Kanpur 208016}

\begin{abstract}
      We develop a semiclassical theory for electron wavepacket dynamics in the presence of an inhomogeneous AC electric field. While static electric-field gradients are known to generate charge transport governed by the quantum metric, we show that AC field gradients induce an additional geometric current that vanishes in the DC limit. This response originates from a novel band-geometric quantity, the higher-order connection (HOC) tensor, constructed from cubic products of interband Berry connections. We derive explicit expressions for the AC current and identify the symmetry conditions under which it arises. Remarkably, inhomogeneous AC fields can generate an anomalous Hall-like response even in nonmagnetic systems. Applying the theory to Bernal-stacked bilayer graphene, we demonstrate that the HOC-induced response produces a measurable Hall current peaking at band edges. These results establish inhomogeneous AC fields as a powerful probe of higher-order band geometric quantities beyond Berry curvature and the quantum metric. 
\end{abstract}
\maketitle

\section{Introduction}
The semiclassical dynamics of electron wavepackets in solids has provided a powerful framework for uncovering the role of band geometry in quantum transport and optical phenomena~\cite{Kohn1999,sundaram_PRB1999_wave,xiao_RMP2010_berry,Cayssol_2021,Adak2024,bohm2013geometric,Sinha2022}. In particular, the anomalous velocity arising from Berry curvature underlies the quantized Hall response of Chern insulators and the intrinsic anomalous Hall effect in metals. Extensions of the semiclassical equations of motion (EOM), combined with Boltzmann theory, have also proven indispensable in identifying geometric contributions to magneto-transport and nonlinear responses~\cite{sodemann_PRL2015_quan,Das2019_PRB,Das_PRR_2020,Ozawa2021,Michishita2022,Mandal2022_PRB,Zeng2021_prb,Fang2024_prl}. However, the conventional semiclassical framework has largely been developed for Bloch electrons in spatially uniform external fields~\cite{Eggarter1972,ashcroft_book1976_solid,richter2000semiclassical,Sinitsyn_2008}. More recently, this framework has been generalized to account for spatially nonuniform electric fields, revealing that the quantum metric governs novel transport responses in metals under static field gradients~\cite{gao_PRL2019_non,Lapa2019,Kozii2021}.  

Despite these advances, a comprehensive framework for electron dynamics under inhomogeneous AC electric fields is still lacking. In particular, the role of band geometry in such AC gradient-driven regimes is not fully understood, even at the level of linear response~\cite{jia_PRB2024_equiv}. This gap presents significant opportunities to uncover and characterize quantum geometric effects across a broader class of transport and optical phenomena.

Here, we generalize the semiclassical EOM of Bloch electrons to account for inhomogeneous AC electric fields with finite spatial gradients. In contrast to the DC case, where corrections are governed by the quantum metric and its derivatives, we find that oscillatory gradient fields induce an additional geometric contribution to the current. This contribution serves as a probe of a unique third-rank band geometric quantity, the higher-order connection (HOC) tensor. 
Remarkably, HOC-induced charge transport produces a measurable anomalous Hall like response even in nonmagnetic systems, provided the system intrinsically breaks inversion symmetry. The HOC tensor is composed of cubic products of Berry connections and it opens a new channel for band-geometry-driven transport, complementing Berry curvature and the quantum metric. As a concrete example, we apply our theory to Bernal-stacked bilayer graphene (BLG) \cite{Sun2012, Kadi2014,Liu2014,Bittencourt2017,Datta2024_nl,Ahmed2025_Small}, demonstrating an anomalous Hall like charge current. These results establish AC inhomogeneous field–induced transport as a sensitive probe of higher-rank band geometric quantities and motivate further exploration of valley and spin transport. 

The paper is organized as follows. In Sec.~\ref{section_two}, we formulate the semiclassical EOM for AC fields with spatial gradients using the wavepacket approach. Sec.~\ref{section_three} derives explicit expressions for the linear current response and highlights the emergence of anomalous Hall–like transport in nonmagnetic systems. In Sec.~\ref{section_four}, we apply the framework to a tight-binding model of Bernal-stacked BLG. We conclude the paper in Sec.~\ref{Section_six} with a summary and outlook. 

\section{AC semiclassical theory in Inhomogeneous Fields}\label{section_two}

We now construct the semiclassical EOM for Bloch electrons under inhomogeneous AC electric fields, beginning with the wavepacket approach. While the EOM describing the dynamics of the electronic wavepacket have been derived for homogeneous AC fields and inhomogeneous DC fields, the influence of an AC inhomogeneous electric field on the EOMs have not been explored earlier. 
Here, we generalize the EOMs and identify a novel band-geometric contribution to the wavepacket velocity,  arising from the electric field gradient term. This velocity contribution, featuring a higher-order band geometric quantity involving the product of three Berry connections, is finite only at nonzero frequencies and vanishes in the DC limit.

\subsection{Wave-Packet Dynamics}

We begin with a perturbative treatment of the Bloch electron wavepacket to understand its dynamics.
In this, we treat the inhomogeneous electric field to lowest order by retaining terms proportional to the first order in field strength. The Hamiltonian for a crystalline material interacting with an external electric field is expressed as \cite{aversa_PRB1995_nonlin, Sipe99},
\begin{equation}    
\hat{\mathcal{H}} = \hat{\mathcal{H}}_0 -e  \hat{\phi}(\hat{{\bm{r}}})~.
\end{equation}
Here, $\hat{\mathcal{H}}_0 = (1/2m)\delta^{\mu \nu}\hat{p}_\mu\hat{p}_\nu + \hat{V}(\hat{{\bm{r}}})$ describes a Bloch electron of mass $m$ in a periodic potential $\hat{V}(\hat{{\bm{r}}})$ with position $\hat{\bm r}$, and momentum $\hat{\bm p}$ operators satisfying 
$[\hat{r}^\mu,\hat{p}_\nu] = i\hbar \delta^\mu_\nu$. We introduce the field-matter interaction using a length gauge approach  \cite{Lapa2019, Kozii2021}, with the scalar potential $\hat{\phi}(\hat{{\bm{r}}})   = - E_\mu \hat{r}^\mu - (1/2) E_{\mu \nu}\hat{r}^\mu \hat{r}^\nu$ expanded perturbatively in the weak field approximation. The effective electric field is given by $\tilde{E}'_\mu= - \partial_{r^\mu }\phi({\bm{r}}) = \tilde{E}_\mu + \tilde{E}_{\mu \nu}r^\nu$. It has a uniform field contribution $\tilde{E}_\mu =  E_\mu e^{-i\omega_1 t}$, and an inhomogeneous gradient term, $\tilde{E}_{\mu \nu} = E_{\mu \nu} e^{-i\omega_1 t}$, where $E_\mu$ and ${E}_{\mu\nu}$ are symmetric tensors, and $\omega_1=\pm\omega$ is the oscillation frequency. For brevity, summations over frequency indices are suppressed and assumed throughout, unless explicitly shown.
  
To probe the dynamics, we construct a wavepacket centered at momentum $\bm{k}_c$ in band $n$ as \cite{xiao_RMP2010_berry, gao_PRL2014_field, jia_PRB2024_equiv}, 
\begin{equation} 
|W\rangle = \int_{\bm{k}} e^{i\bm{k} \cdot \bm{r}} \left[ C_n(\bm{k}) |u_n\rangle + \sum_{l \neq n} \bar{C}_l(\bm{k}) |u_l\rangle \right]~. \label{wavepacket} 
\end{equation} 
Here, $|u_n\rangle \equiv |u_n(\bm{k})\rangle$ is the cell-periodic part of the Bloch function and $\int_{\bm{k}} \equiv \int d\bm{k}/(2\pi)^D$ in $D$ dimensions. The coefficient $|C_n(\bm{k})|^2 \simeq \delta(\bm{k} - \bm{k}_c)$ defines the momentum of the center of the wavepacket and $\bar{C}_l(\bm{k})$ captures the interband mixing induced by the electric field \cite{xiao_RMP2010_berry}. The wavepacket evolves via the time-dependent Schr\"odinger equation (TDSE), \begin{equation}
    i\hbar \partial_t |W\rangle = \hat{\mathcal{H}} |W\rangle~.\label{tdse}
\end{equation}
The unperturbed Bloch basis states are time-independent with $\langle u_n | \partial_t u_n \rangle = 0$ \cite{xiao_RMP2010_berry}. 
To ensure well-defined solutions under time-dependent perturbation, we employ adiabatic switching approximation and use a positive infinitesimal damping parameter $\xi \to 0^+$ \cite{Kumar2024, Bhalla2022,sarkar2025_3rdrect} to modify the electric field $\tilde{E}'_\mu =  E_\mu e^{-i(\omega_1 + i \xi)t} +  E_{\mu\nu} e^{-i(\omega_1 + i \xi)t}r^\nu$. We use this to calculate the perturbative correction to the wavepacket up to linear order in the applied field strength. This resulting wavepacket is given by,  
\begin{align}
|W\rangle
&=
\int_{\vect{k}} e^{i\vect{k}\cdot\vect{r}}
C_{n} (\vect{k})
\left[
|u_{n}\rangle
+
\sum_{m \neq n}  M_{mn}^{(1)} |u_{m} \rangle
\right]~,
\nonumber \\
&\equiv
|W_0\rangle+|W_1\rangle~.
\label{wavepacket1}
\end{align}
The derivation and details of the coefficient $M_{mn}^{(1)}$ are presented in Appendix ~\ref{App_A}. This perturbed wavepacket serves as the foundation for deriving the EOMs for the wavepacket center, which reveals previously unexplored band geometric terms arising from the inhomogeneity and finite-frequency effects of the applied field. 

\begin{table*}[t]
	\centering
     \caption{This table summarizes contributions to the linear transport currents induced by spatially varying, oscillating electric fields. Each contribution is expressed in terms of band geometric quantities such as  Berry curvature ($\Omega_{nm}^{\mu\nu}$), quantum metric ($\mathcal{G}_{nm}^{\mu\nu}$), and the higher-order connection (HOC) defined as $\mathcal{N}^{\nu\mu\lambda}_{nmp} = \mathcal{R}^\nu_{mn} \mathcal{R}^\mu_{np} \mathcal{R}^\lambda_{pm}$. Here, $F_{mn} = f_m - f_n$ denotes the difference in occupancy of bands $m$ and $n$ and $\partial_\mu\equiv\frac{\partial}{\partial k_{\mu}}$. The last four columns indicate the leading symmetry-allowed contributions under parity or inversion ($\mathcal{P}$) and time-reversal ($\mathcal{T}$) symmetry ($\mathcal{P}\checkmark\mathcal{T}\checkmark$), $\mathcal{P}$ in absence of $\mathcal{T}$ ($\mathcal{P}\checkmark\mathcal{T}\times$), $\mathcal{T}$ in absence of $\mathcal{P}$ ($\mathcal{P}\times\mathcal{T}\checkmark$), and combined $\mathcal{PT}$ symmetry. Notably, different symmetries selectively allow or constrain specific geometric responses, offering a tunable parameter for exploring selective responses.                                                          
 }
	{
		\begin{tabular}{c c c c c c}
			\hline \hline
			\rule{0pt}{3ex}
			  Current & Integrand ~~&~~$\mathcal{P}\checkmark\mathcal{T}\checkmark$ ~~&~~ $\mathcal{P}\checkmark\mathcal{T}\times$ ~~&~~ $\mathcal{P}\times\mathcal{T}\checkmark$ &   ${\mathcal P}{\mathcal T}$ \\ [1ex]
			\hline \hline
			\rule{0pt}{3ex}
			 $j_{\mu}^{\rm Dr}$ & $-\frac{ie^2}{\hbar^2}(\tilde{E}_\nu + \tilde{E}_{\nu \lambda} r^\lambda)\sum_{n}\dfrac{\partial_{\mu} \epsilon_n  }{(\omega_1+i/\tau)} ~\partial_{\nu} f_n$ & $\neq 0$& $\neq 0$ &$\neq 0$ &  $\neq 0$  \\ [2ex]
			 $j_{\mu}^{\rm An}$ & $ \frac{e^2}{\hbar}(\tilde{E}_\nu + \tilde{E}_{\nu \lambda} r^\lambda)\sum_{n,m}{\Omega}^{\mu\nu}_{nm}f_n $ & 0 & $\Omega$  &  0  &  0  \\[2ex]
			$j_{\mu}^{\rm QM}$ & 	$-\frac{e^2}{2\hbar}\sum_{n,m}\tilde{E}_{\nu \lambda}{\partial_\mu \mathcal{G}^{\nu\lambda }_{mn}} f_n$ & 0 & 0 & 0 & $\mathcal{G}$ \\ [2ex]
			$j_{\mu}^{\rm QGT}$ &  $ \frac{ie^2}{\hbar}\tilde{E}_{\nu}\sum_{n,m}\bigg[ \omega_1\bigg(\mathcal{G}^{\nu \mu}_{nm} - \frac{i}{2} \Omega^{\nu \mu}_{nm} \bigg) g^{\omega_1}_{mn} \bigg] F_{nm}$ & $\mathcal{G}$ & $\mathcal{G}$, $\Omega$ & $\cal{G}$ & $\mathcal{G}$ \\[2ex]
			 $j_{\mu}^{\rm HOC}$ ~~&~~ $\frac{ie^2}{2\hbar}\tilde{E}_{\lambda \nu}  \sum_{ n,m,p} \omega_1\left( \mathcal{N}_{nmp}^{\mu\nu\lambda}f_n- \mathcal{N}_{nmp}^{\mu\lambda\nu}f_m\right)g^{\omega_1}_{mn}$  & 0 & 0 & ${\rm Re}[\mathcal{N}]$ ~&~ ${\rm Im}[\mathcal{N}]$  \\[2ex]		
						\hline \hline
	\end{tabular}}
	\label{table_1}
\end{table*}
\subsection{Equations of motion}

We first calculate the position of the electron wavepacket in real space, which is defined as $r^\mu_c = \langle W| \hat{r}^\mu|W\rangle$. This will be used to calculate the Lagrangian of the system and the semiclassical EOMs. We obtain \cite{gao_PRL2014_field}, 
\begin{equation}
    r^\mu_c
=
\partial_{k_{c,\mu}}\gamma  + \mathcal{R}^\mu_{nn}  + a_n^{\mu}~, 
\end{equation}
where $\gamma$ is the phase factor in the complex coefficient $C_n(\vect k_c) = |C_n(\vect k_c)| e^{-i\gamma(\vect k_c)}$, $\mathcal{R}_{nm}^\mu=i\langle u_n|\partial_{k_{\mu}} u_m\rangle$ is the $\mu$th component of the interband Berry connection, and
 $a_n^{\mu}$ denotes the  \textit{positional shift} of the wavepacket due to the electric field perturbation. This field-induced shift captures the effects of both spatially homogeneous and inhomogeneous components of the AC electric field. It takes the general form,
\bea \label{an_mu_gen}
a_n^{\mu}
&=& a_n^{\mu,{\rm hom}} + a_n^{\mu,{\rm inhom}}~, \
\eea  
where, taken up to the first order correction,  $a_n^{\mu}$ is given as, 
\bea
a_n^{\mu} \equiv 2{\rm Re} \langle W_0|r^\mu|W_1\rangle~ =~ 
2{\rm Re} \sum_{m}\left[
M_{mn}^{(1)}
\mathcal{R}^\mu_{nm}\right]~.
\eea
The first term in Eq. \eqref{an_mu_gen}, $a_n^{\mu,{\rm hom}}$ captures the shift induced by a spatially uniform AC field, while $a_n^{\mu,{\rm inhom}}$ accounts for the shift by the spatial gradient. We obtain, 
\begin{equation}
a_n^{\mu,{\rm hom}}=2\rm{Re}\sum_m\left[ \tilde{E}_\nu \mathcal{Q}^{\mu\nu}_{\it mn}  g^{\omega_1}_{\it mn}~\right]~. 
\label{eq_hom}
\end{equation}
Here, we have defined $g^{\omega_1}_{mn} = [(\omega_1 - \omega_{mn}) + i/\tau]^{-1}$, with $1/\tau = \xi$. Finite $\tau$ in $g^{\omega_1}_{mn}$ takes care of the divergences appearing in the theory in the $\omega \to 0$ regime \cite{kaplan_Sci_uni,Xiang2024,jia_PRB2024_equiv}. In the above equation, the quantum geometric tensor (QGT) \cite{ahn_NP2022_rieman,cheng2013qgt_fsm,Bhalla2022,bhalla_PRB2023_quan} is defined as,
\bea \label{QGT_eq}
\mathcal{Q}_{mn}^{\mu\nu} &=& \mathcal{R}_{nm}^\mu \mathcal{R}_{mn}^\nu = \mathcal{G}_{mn}^{\mu\nu} - \frac{i}{2} \Omega_{mn}^{\mu\nu}~,
\eea
where $\mathcal{G}_{mn}^{\mu\nu}$ and $\Omega_{mn}^{\mu\nu}$ are the band-resolved quantum metric and Berry curvature, respectively.
Eq.~\eqref{eq_hom} reproduces the linear-order shift in the center of the wavepacket identified in  Ref.~\cite{gao_PRL2014_field}. This linear positional shift gives rise to the first-order correction in the EOMs driven by a homogeneous AC electric field \cite{gao_PRL2014_field,jia_PRB2024_equiv}. 

The wavepacket center shift induced by an AC inhomogeneous field is given by, 
\begin{equation}
 a_n^{\mu,{\rm inhom}} =\rm{Re}\sum_{\it m,p}\left[{\tilde{E}_{\nu\lambda}}   \mathcal{N}^{\mu\nu\lambda}_{\it nmp}  g^{\omega_1}_{\it mn}\right]~.
 \label{eq_inhom}
\end{equation}
Here, we have defined the HOC - a new gauge-invariant band geometric quantity, as, 
\bea
\mathcal{N}^{\mu\nu\lambda}_{nmp}=\mathcal{R}^\mu_{nm} \mathcal{R}^\nu_{mp}  \mathcal{R}^\lambda_{pn}={\rm Re}[\mathcal{N}^{\mu\nu\lambda}_{nmp}]+i{\rm Im}[\mathcal{N}^{\mu\nu\lambda}_{nmp}]~.\nn\\
\eea 
This term in Eq.~\eqref{eq_inhom} has not been explored earlier, and it enables the generalization of the semiclassical wavepacket dynamics to include the impact of AC field gradient.

\begin{figure*}[t!] 
    \includegraphics[width=1\linewidth]{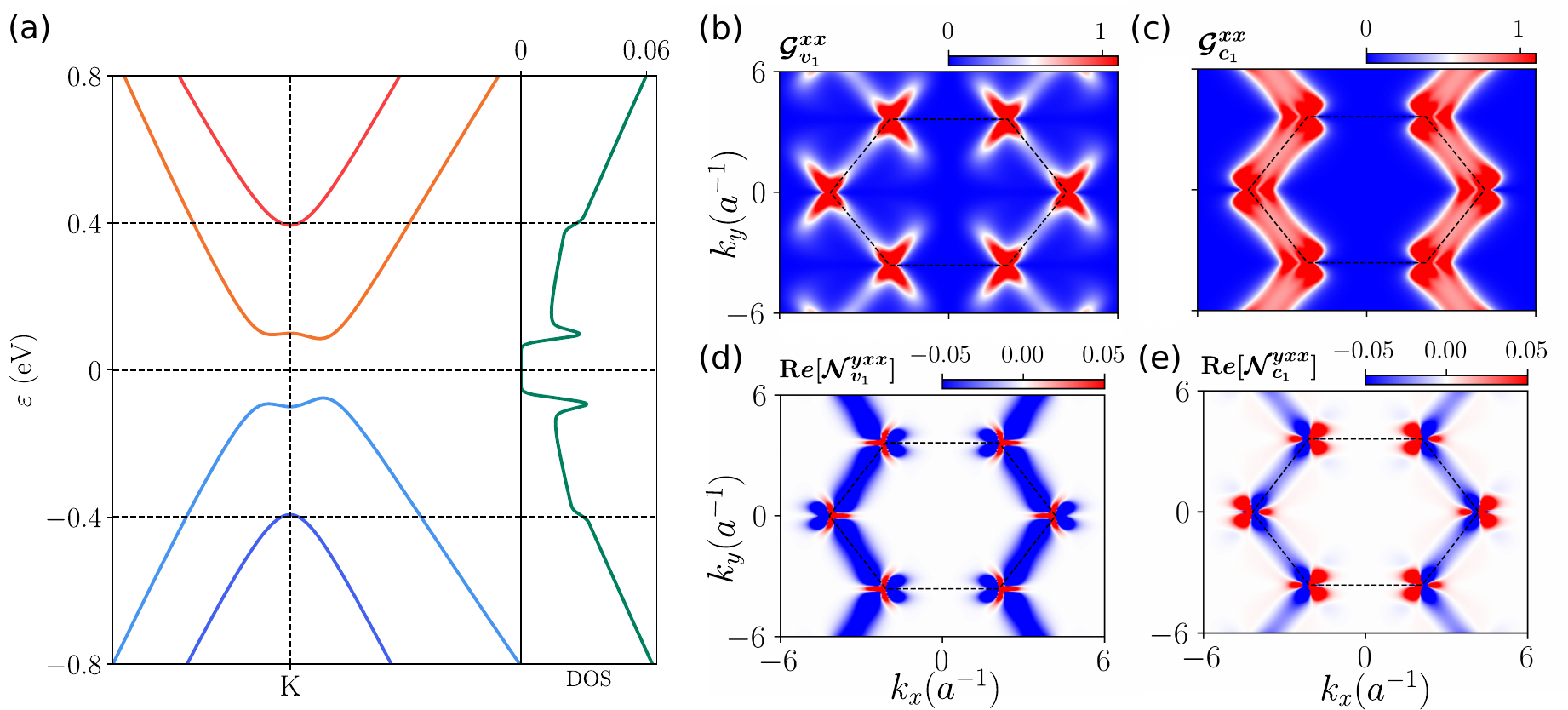}
    \caption{ (a) The electronic band structure of BLG around the $K$ point. The corresponding density of states is shown alongside in units of $a^{-2}{\rm eV}^{-1}$. Panels (b) and (d) present the momentum-space distribution of the quantum metric component $\mathcal{G}^{xx}$ and the real part of the HOC, ${\rm Re}[\mathcal{N}^{yxx}]$, respectively, for the upper valence band.
Panels (c) and (e) show the corresponding quantities, $\mathcal{G}^{xx}$ and ${\rm Re}[\mathcal{N}^{yxx}]$, for the lower conduction band. We have set the interlayer potential to be $\Delta=0.1$ eV.
\label{Fig1}}
\end{figure*}

To derive the semiclassical EOMs, we construct the effective Lagrangian incorporating the field-induced corrections to the wavepacket. Following Ref.~\cite{jia_PRB2024_equiv}, we obtain, 
\begin{equation}
    \mathcal{L}
    = -\dot{\bm{k}}_{c}  \cdot (\bm{r}_c - \bar{{\vect{\mathcal{R}}}}_{nn}) - \tilde{\epsilon}_n~. \label{L_WD}
\end{equation}
 Here, $\bar{\mathcal{R}}_{nn}^\mu = \mathcal{R}_{nn}^\mu + a^{\mu}_n,$ is the $\mu$th component of the modified Berry connection that includes the first-order field-induced correction, and the corrected energy of the wavepacket, $\tilde{\epsilon}_n,$ is given as \cite{Lapa2019, gao_PRL2019_non},
\begin{eqnarray}
 \tilde{\epsilon}_n &=& \langle{W}_0|i \partial_t| W_0 \rangle - \langle W|i \partial_t - \mathcal{H}| W \rangle ~,\nn\\
 &=& \epsilon_n+e\tilde{E}_\mu r_c^\mu+\frac{e\tilde{E}_{\mu\nu}}{2}\sum_{m\neq n} \mathcal{G}^{\mu\nu}_{mn}~,
\end{eqnarray} 
where the last term captures the energy correction due to the field gradient. This energy correction induced by the AC inhomogeneous electric field has also been shown to give a DC contribution to the effective band velocity ~\cite{Lapa2019}. 
Substituting this into the Lagrangian and using the Euler-Lagrange equation \cite{morin2008introduction}, we obtain the semiclassical EOMs as, 
\begin{widetext}
\bea
    \dot{r}^\mu_{c, n} =\frac{1}{\hbar}\frac{\partial \epsilon_n}{\partial k_{c,\mu}}+ \frac{e}{2\hbar}\sum_{m\neq n}\tilde{E}_{\nu\lambda }\frac{\partial \mathcal{G}^{\nu\lambda }_{mn}}{\partial k_{c, \mu}} - \sum_{m\neq n} \dot{k}_{c, \nu} \Omega_{mn}^{\mu\nu} + 2\rm{Re}~ \bigg\{ ~\partial_{\it t} \bigg[\bigg(\sum_{\it m} \tilde{E}_\nu \mathcal{Q}^{\mu\nu}_{\it mn}\bigg)  g^{\omega_1}_{\it mn} + \frac{\tilde{E}_{\nu\lambda}}{2}  \bigg( \sum_{\it m,p}\mathcal{N}^{\mu\nu\lambda}_{\it nmp} \bigg) g^{\omega_1}_{\it mn}\bigg]\bigg\}~, \nn \\
    \label{EL_eqns1}
\eea    
\bea
    \hbar\dot{k}_{c, \mu}=-e\left(\tilde{E}_\mu + \tilde{E}_{\mu\nu}r^\nu_c\right)~.\label{EL_eqns2}
\eea
\end{widetext}

In the above expression for band velocity, the Berry curvature of the $n$th band can be expressed as a sum over interband contributions, i.e., $\Omega_{n}^{\mu\nu} = \sum_{m\neq n} \Omega_{mn}^{\mu\nu}$. Hereafter, we suppress the subscript $c$ for brevity. Without inhomogeneity ($\tilde{E}_{\mu\nu}=0$), Eqs.~\eqref{EL_eqns1} and~\eqref{EL_eqns2} recover the first-order AC semiclassical theory of Sundaram and Niu ~\cite{sundaram_PRB1999_wave}. The  $\partial_t a_n^{\mu}$ term contributes only at finite frequencies and yields displacement currents~\cite{Xiang2024}. The second term in $\dot{r}^\mu_{c}$ in Eq.~\eqref{EL_eqns2}, which involves the electric-field gradient and the quantum metric, leads to quantum-metric driven current, identified in Ref.~\cite{Lapa2019}. These EOMs, combined with the equilibrium distribution function, generate novel AC currents under an inhomogeneous electric field. We explore this below.

\section{Linear responses}\label{section_three}
Building on the semiclassical EOMs derived in the previous section, we now calculate the linear current response at position $\bm{R}$, under a spatially inhomogeneous AC electric field.  The local current density is given by \cite{Lapa2019}, 
\be
j^\mu (\bm{R}) = -e\sum_n\int_{\bm{r}}\int_{\bm{k}}\dot{r}_n^\mu \bar{f}_n\delta^D(\bm{r} - \bm{R})~. \label{current}
\ee
Here, $\int_{\bm{r}}$ is the $D$-dimensional real space integral, and $\bar{f}_n$ is the nonequilibrium distribution function (see Appendix~\ref{App1}) describing the occupancy of the $n$th band electrons in the infinitesimal phase-space volume  ${d\bm{r} ~d\bm{k}}/{(2\pi)^D}$. Substituting the EOM from Eq.~\eqref{EL_eqns1} and $\bar{f}_n$, we identify five distinct contributions to the linear current originating from band velocity and band geometry: (i) Drude current, (ii) anomalous current, (iii) quantum metric induced current, (iv) QGT-induced current, and (v) HOC-induced current. Table~\ref{table_1} summarizes these responses along with their expressions and fundamental symmetry constraints. Notably, the last two terms vanish in the DC limit.
The HOC-induced current offers a new transport mechanism for probing band geometric properties of quantum matter with inhomogeneous AC fields. 

The Drude current, $j^{\rm{Dr}}_{\mu},$ arises from intraband transitions and captures the conventional band velocity contributions under both uniform and gradient electric fields \cite{Smith2001}. The anomalous current, $j^{\rm{An}}_{\mu}$, originates from the band-resolved Berry curvature coupled to the effective electric field \cite{Nagaosa2010_rmp,Sinitsyn2007, Culcer_2024, Jungwirth2002, Fang2003_sci}. It vanishes in time-reversal ($\mathcal{T}$) symmetric systems (where Berry curvature is antisymmetric under $\mathcal{T}$) and in $\mathcal{PT}$-symmetric antiferromagnets (where Berry curvature is zero throughout the Brillouin zone). 
\begin{figure*}[t!] 
    \includegraphics[width=\linewidth]{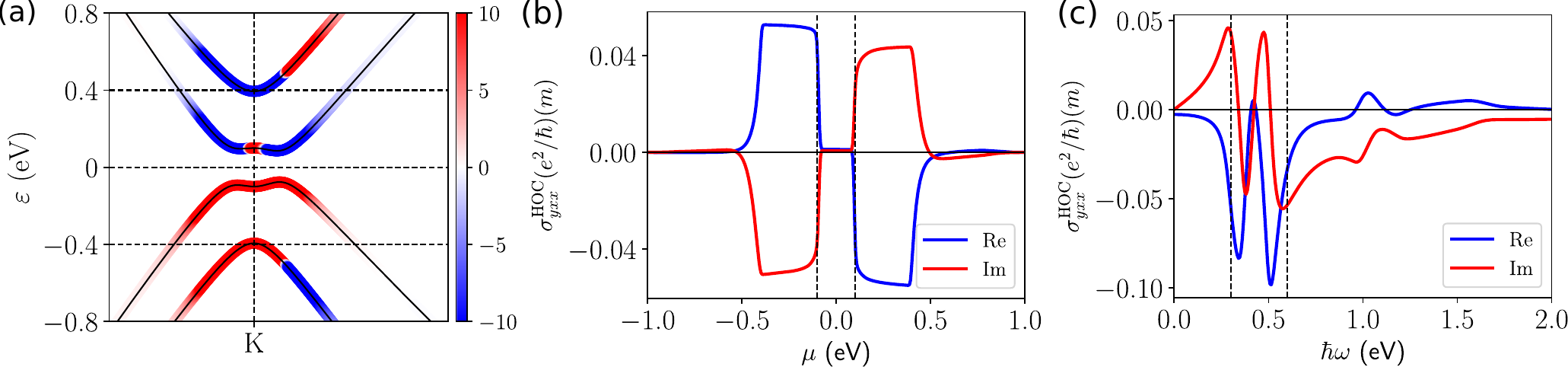}
    \caption{(a) Band structure of bilayer graphene near the $K$ point, with the color scale indicating the higher-order connection tensor. 
    The $yxx$-component of the anomalous Hall-like conductivity in bilayer graphene, induced by the higher-order connection tensor, as a function of (b) chemical potential $\mu$  for photon energy $\hbar\omega = 0.3$ eV, and (c) photon energy $\hbar\omega$ for a fixed $\mu$. The real and imaginary parts of the conductivity are shown in blue and red, respectively. In panel (b), the dashed lines mark the band edges. In panel (c), the vertical dashed lines mark $\hbar\omega = \mu$ and $\hbar\omega = 2\mu$, which correspond to key resonant features.  We use a band gap $\Delta = 0.1$ eV and a relaxation time $\tau = 1.3 \times 10^{-14}$ s, and temperature $T = 30$ K.
\label{Fig2}}
\end{figure*}
The quantum metric-driven current, $j^{\rm{QM}}_{\mu},$ emerges only under an inhomogeneous electric field and is geometrically linked to the momentum-space gradient of the quantum metric \cite{Das2023_qmt, Gao2023}. Both QGT-induced and HOC-induced currents, $j^{\rm{QGT}}_{\mu}$ and $j^{\rm{HOC}}_{\mu}$, originate from the time derivative of the first-order field-induced positional shift captured by $a_n^{\mu}$ (see Appendix~\ref{App2} for details). The QGT term responds to the total field \cite{Kang2025, Bleu2018}, while the HOC contribution depends solely on the gradient of the field and it vanishes in the uniform-field limit. Except for the Drude response, which arises from Fermi surface effects, all other current responses involve Fermi sea carriers \cite{Hodges1971,bhalla_PRB2023_quan}.  

The HOC response arises from a higher-order band geometric structure involving a triple product of the Berry connection. It vanishes under inversion symmetry and exhibits distinct parity: only $\rm{Re}[\mathcal{N}]$  contributes in $\mathcal{T}$ symmetric materials, and only $\rm{Im}[\mathcal{N}]$ under combined $\mathcal{PT}$ symmetric antiferromagnets. Remarkably, in $\mathcal{T}$ symmetric nonmagnetic materials, the HOC is the sole gradient field-dependent charge response. This previously unexplored geometric contribution to charge response is the central contribution of this work. 

\section{Linear Current in Bilayer Graphene}\label{section_four}
To demonstrate the linear current induced by an AC inhomogeneous electric field, we study the Bernal-stacked BLG  \cite{McCann_2013, MCCANN2007, Ohta2006, Yan2015, Butz2014,Ghorai2025}, within a tight-binding framework (see Appendix \ref{App_C} for details). The BLG is intrinsically a $\cal P$ and $\cal T$ symmetric system, but its $\cal P$ symmetry can be easily broken by applying a vertical electric field in a dual-gated device geometry. As highlighted in Table~\ref{table_1}, $\cal P$ broken $T$ preserving systems can support the HOC current induced by ${\rm Re} {[\cal N]}$, while they suppress the corresponding anomalous and quantum metric induced contributions. 

In its pristine form, the BLG respects $D_{3d}$ point group symmetry, which includes inversion symmetry. However, when a perpendicular (out-of-plane) electric field is applied, the inversion symmetry is broken, reducing the symmetry from $D_{3d}$ to $C_{3z}$ and opening a tunable band gap. The resulting Hamiltonian with a vertical electric field preserves time-reversal symmetry and threefold rotational symmetry. 

We present the band structure along with the density of states (DOS) of BLG around the K point in Fig.~\ref{Fig1}(a). Panels (b)–(e) depict the momentum-space distribution of the quantum metric and the real part of the HOC tensor in the conduction and valence band, highlighting their significant values near the band edges or along the lines connecting the $K$ and $K'$ points in the Brillouin zone. Fig.~\ref{Fig2}(a) further illustrates the HOC weight on the BLG bands in the $K$ valley. 

The HOC-induced current can also be expressed in terms of third-rank conductivity tensors. For example, the current induced in the $y$-direction due to an electric field gradient along the $x$-direction is $j_y^{\rm HOC}=\sigma_{yxx}^{\rm HOC}\tilde{E}_{xx} $. The $C_{3v}$ symmetry constraint in BLG allows four in-plane tensor components of the HOC conductivity to be finite. However, only one of these components is independent. The components are related as
\be
\sigma_{yxx}^{\rm HOC}=-\sigma_{yyy}^{\rm HOC}=\sigma_{xxy}^{\rm HOC}=\sigma_{xyx}^{\rm HOC}~. 
\ee
 
We numerically evaluate the real and imaginary parts of $\sigma_{yxx}^{\rm HOC}$, and show their variation with chemical potential $\mu$, and the optical frequency $\omega$ in Figs.~\ref{Fig2}(b) and \ref{Fig2}(c), respectively. The dominant contributions arise from the primary conduction and valence bands near the bandgap, consistent with the HOC distribution in Fig.~\ref{Fig2}(a). 
As the chemical potential enters the conduction or valence band, the conductivity becomes finite and remains approximately constant up to $|\mu| \approx 0.4$ eV, where only a single band contributes. Beyond this, the onset of higher bands reduces the HOC conductivity despite the increasing density of states in the system [see Fig.~\ref{Fig1}(b)]. Thus, HOC-dominated regions near the band edges primarily drive the anomalous Hall–like responses, which can be sensitively tuned through band filling and vertical electric fields in BLG.

\section{Conclusion}\label{Section_six}

Inhomogeneous electric fields offer a promising avenue to probe the quantum geometric structure of Bloch bands beyond conventional semiclassical treatments. In this work, we developed a semiclassical framework to describe electron dynamics in the presence of spatially varying, AC electric fields. Notably, we identified two distinct contributions to the quasiparticle velocity: one originating from the quantum geometric tensor, and the other from the higher-order connection tensor. Both effects emerge naturally and persist even in the linear response regime. 
Among these, the higher-order-connection induced response is particularly striking, as it arises from the interplay between the gradient component of the electric field and its finite frequency. Remarkably, the higher-order-connection induced contribution gives rise to an anomalous Hall like response even in nonmagnetic systems, as long as they intrinsically break inversion symmetry. We demonstrate this anomalous Hall-like transverse current in pristine Bernal-stacked bilayer graphene (BLG) with broken inversion symmetry. Our findings establish AC inhomogeneous field–induced transport as a versatile probe of band geometry in crystalline solids and provide a foundation for exploring spin, valley, and nonlinear charge transport in future studies.

\section{Acknowledgment}
S.S. thanks Harsh Varshney (IIT Kanpur) for the useful discussion. 
S. S. thanks the MHRD, India, for funding through the Prime Minister’s Research Fellowship (PMRF). A. A. acknowledges funding from the Core Research Grant by Anusandhan National Research Foundation (ANRF, Sanction No. CRG/2023/007003).

\appendix

\section{Calculation of $M_{mn}^{(1)}$}\label{App_A}
\begin{widetext}
In this Appendix, we present the calculation of the coefficient $M^{(1)}_{mn}$, which is needed to obtain the first-order correction to the wavepacket. For this, we start with the TDSE in Eq.  \eqref{tdse}, and multiply on both sides with $\langle u_m| e^{-i \bm{k}\cdot \bm{r}}$. This leads to the following equation,  
\begin{align}
&{\hbar} \int_{\vect{k}'} e^{i(\vect{k}'-\vect{k})\cdot\vect{r}}
\left(
i\dot{C}_n(\vect{k}')\langle u_m|u_n' \rangle
+
\sum_{l \neq n}
i \dot{\bar{C}}_l(\vect{k}') \langle u_m|u_l' \rangle
\right)
=
\int_{\vect{k}'}
e^{i(\vect{k}'-\vect{k})\cdot\vect{r}}
\left(
C_n(\vect{k}')\langle u_m|H_0|u_n' \rangle
+
\sum_{l \neq n}
\bar{C}_l(\vect{k}') \langle u_m|H_0 |u_l'\rangle
\right)
\nonumber \\
& \qquad \qquad \qquad \qquad \qquad \qquad \qquad \qquad
+
\int_{\vect{k}'}
e^{i(\vect{k}'-\vect{k})\cdot\vect{r}}
\left(
C_n(\vect{k}')\langle u_m|H_1|u_n' \rangle
+
\sum_{l \neq n}
\bar{C}_l(\vect{k}')
\langle u_m|H_1|u_l' \rangle
\right)~.
\label{SE1}
\end{align}
We simplify the obtained expression using the following identities \cite{ aversa_PRB1995_nonlin, Sipe99,Lapa2019, jia_PRB2024_equiv}, 
\begin{eqnarray}
\langle u_m| e^{i(\bm{k}' - \bm{k})\cdot \bm{r}} | u_n \rangle &=& \delta(\bm{k}' - \bm{k}) \delta_{mn}~, \\    
\langle u_m| e^{i(\bm{k}' - \bm{k})\cdot \bm{r}} r^\mu| u_n \rangle &=& [\delta_{mn} i \partial_{k_\mu} + \mathcal{R}^\mu_{mn}]\delta(\bm{k}' - \bm{k})~, \\   
\langle u_m| e^{i(\bm{k}' - \bm{k})\cdot \bm{r}} r^\mu r^\nu| u_n \rangle &=& \sum_p[-\delta_{mp}\delta_{pn} \partial_{k_\mu} \partial_{k_\nu}+ i\mathcal{R}^\mu_{mp}\partial_{k_\nu}\delta_{pn}+i\mathcal{R}^\nu_{pn}\partial_{k_\mu}\delta_{mp}+\mathcal{R}^\mu_{mp}\mathcal{R}^\nu_{pn}]\delta(\bm{k}' - \bm{k})~,    
\end{eqnarray}
where, $\partial_{k_\mu} \equiv \partial/ \partial k_\mu,$ and $\mathcal{R}^\mu_{mn} = \langle u_m | i \partial_{k_\mu} u_n \rangle$ is the interband Berry connection. For $m=n$, Eq. \eqref{SE1} becomes, 

\begin{eqnarray}
i {\hbar} \dot{C}_n  &=& {\hbar}\omega_n C_n
+
{e}\tilde{E}_\mu \left( i \partial_{k_\mu} + \mathcal{R}_{nn}^\mu \right) C_n + {e}\frac{\tilde{E}_{\mu \nu}}{2} \left( -\partial_{k_\mu}\partial_{k_\nu} + i\mathcal{R}_{nn}^\mu\partial_{k_\nu} + i\mathcal{R}_{nn}^\nu\partial_{k_\mu} +\sum_p\mathcal{R}_{np}^\mu\mathcal{R}_{pn}^\nu\right)C_n
+ {e}\sum_{l} \bar{C}_l \tilde{E}_\mu  \mathcal{R}^\mu_{nl} 
\nn\\
& {} & + {e}\frac{\tilde{E}_{\mu \nu}}{2}\sum_l \bar{C}_l \left( i\mathcal{R}^\mu_{nl}\partial_{k_\nu}+i\mathcal{R}^\nu_{nl}\partial_{k_\mu}+\sum_p \mathcal{R}^\mu_{np}\mathcal{R}^\nu_{pl}\right)~.
\label{Cn}
\end{eqnarray} 
Similarly, when $m \neq n$ we get, 
\begin{eqnarray}
i {\hbar} \dot{\bar{C}}_m &=& {\hbar}\omega_m \bar{C}_m
+
{e}\tilde{E}_\mu\mathcal{R}^\mu_{mn} C_n
+{e}\frac{\tilde{E}_{\mu \nu}}{2}\left( i\mathcal{R}^\mu_{mn}\partial_{k_\nu}+i\mathcal{R}^\nu_{mn}\partial_{k_\mu}+\sum_p \mathcal{R}^\mu_{mp}\mathcal{R}^\nu_{pn}\right)C_n +  
{e}\tilde{E}_\mu  \left( i \partial_{k_\mu} + \mathcal{R}^\mu_{mm}\right) \bar{C}_m \nn\\
& {} & +
{e}\sum_{l} \bar{C}_l \tilde{E}_\mu\mathcal{R}_{ml}^\mu +{e}\frac{\tilde{E}_{\mu \nu}}{2} \left( -\partial_{k_\mu}\partial_{k_\nu} + i\mathcal{R}_{mm}^\mu\partial_{k_\nu} + i\mathcal{R}_{mm}^\nu\partial_{k_\mu}+\sum_p\mathcal{R}_{mp}^\mu\mathcal{R}_{pm}^\nu\right)\bar{C}_m \nn\\
& {} & + {e}\frac{\tilde{E}_{\mu\nu}}{2}\sum_{l} \bar{C}_l  \left( i\mathcal{R}^\mu_{ml}\partial_{k_\nu}+i\mathcal{R}^\nu_{ml}\partial_{k_\mu}+\sum_p \mathcal{R}^\mu_{mp}\mathcal{R}^\nu_{pl}\right)~.
\label{barCm}
\end{eqnarray}
Equation \eqref{barCm}, together with Eq. \eqref{Cn}, completely determine the unknown coefficients $\bar{C}_l$
in the wavepacket defined by Eq. \eqref{wavepacket}.
Parametrizing, $\bar{C}_m=M_{mn}C_n$, we find that
Eq.\eqref{barCm} becomes,
\begin{align}
i \hbar (\dot{M}_{mn} C_n + M_{mn} \dot{C}n) &= \hbar \omega_m M_{mn} C_n + e \tilde{E}_\mu \mathcal{R}^\mu_{mn} C_n + e \frac{\tilde{E}_{\mu\nu}}{2} \left( i \mathcal{R}^\mu_{mn} \partial_{k_\nu} + i \mathcal{R}^\nu_{mn} \partial_{k_\mu} + \sum_p \mathcal{R}^\mu_{mp} \mathcal{R}^\nu_{pn} \right) C_n \nonumber \\
&\quad + e \tilde{E}_\mu \left( i \partial{k_\mu} + \mathcal{R}^\mu_{mm} \right) M_{mn} C_n + e \sum_l \tilde{E}_\mu \mathcal{R}^\mu_{ml} M_{ln} C_n \nonumber \\
&\quad + e \frac{\tilde{E}_{\mu \nu}}{2} \left( -\partial{k_\mu} \partial_{k_\nu} + i \mathcal{R}^\mu_{mm} \partial_{k_\nu} + i \mathcal{R}^\nu_{mm} \partial_{k_\mu} + \sum_p \mathcal{R}^\mu_{mp} \mathcal{R}^\nu_{pm} \right) M_{mn} C_n \nonumber \\
&\quad + e \sum_l \frac{\tilde{E}_{\mu \nu}}{2} \left( i \mathcal{R}^\mu{ml} \partial_{k_\nu} + i \mathcal{R}^\nu_{ml} \partial_{k_\mu} + \sum_p \mathcal{R}^\mu_{mp} \mathcal{R}^\nu_{pl} \right) M_{ln} C_n~.
\end{align}

Substituting Eq. \eqref{Cn} into the above expression and eliminating $C_n$, transforms it as, 

\begin{eqnarray}
i{\hbar}\dot{M}_{mn} + {\hbar}\omega_{nm} M_{mn}
&=&
 {e}\tilde{E}_\mu \mathcal{R}_{mn}^\mu
+
i{e}\tilde{E}_\mu\mathcal{D}^{\mu}_{mn} M_{mn} +
{e}\sum_{l}  \tilde{E}_\mu \left(\mathcal{R}_{ml}^\mu-M_{mn}\mathcal{R}_{nl}^\mu\right) M_{ln}+{e}\frac{\tilde{E}_{\mu \nu}}{2}\sum_p\left(\mathcal{R}_{mp}^\mu-\mathcal{R}_{np}^\mu M_{mn}\right)\mathcal{R}_{pn}^\nu \nn\\
&{}& +i{e}\frac{\tilde{E}_{\mu \nu}}{2}\sum_l\left(\mathcal{R}_{ml}^\mu-\mathcal{R}_{nl}^\mu M_{mn}\right)\partial_{k_\nu} M_{ln}+i{e}\frac{\tilde{E}_{\mu \nu}}{2}\sum_l\left(\mathcal{R}_{ml}^\nu-\mathcal{R}_{nl}^\nu M_{mn}\right)\partial_{k_\mu} M_{ln}+{e}\frac{\tilde{E}_{\mu \nu}}{2}\sum_{pl}\left(\mathcal{R}_{mp}^\mu\right.\nn\\
&{}& - \left.\mathcal{R}_{np}^\mu M_{mn}\right)\mathcal{R}_{pl}^\nu M_{ln} + {e}\frac{\tilde{E}_{\mu \nu}}{2}\left(-\partial_{k_\mu}\partial_{k_\nu}+i\mathcal{R}_{mm}^\mu\partial_{k_\nu}+i\mathcal{R}_{mm}^\nu\partial_{k_\mu}+\sum_p\mathcal{R}_{mp}^\mu\mathcal{R}_{pm}^\nu\right)M_{mn}~. \label{Mmn}
\end{eqnarray}
Here, $\mathcal{D}^{\mu}_{mn}=\partial_{k_\mu}-i(\mathcal{R}_{mm}^\mu-\mathcal{R}_{nn}^\mu)$ is the covariant derivative and $\epsilon_{nm}=\epsilon_n-\epsilon_m$ is the energy difference between $n$ and $m$ bands. Equation~\eqref{Mmn} can be solved perturbatively by expressing $M_{mn} = \sum_i M_{mn}^{(i)}$, with $i=0,1,2,...$ and $M_{mn}^{(i)} \propto |\tilde{\bm{E}}'|^i$. For $i=1$, where the correction to $M_{mn}$ is proportional to the linear order in the electric field, we obtain, 
\begin{equation}
i{\hbar}\dot{M}_{mn}^{(1)} +{\hbar}\omega_{nm} M_{mn}^{(1)} =  {e}\left(\tilde{E}_\mu \mathcal{R}^\mu_{mn}+\frac{\tilde{E}_{\mu \nu}}{2} \sum_p\mathcal{R}^\mu_{mp}\mathcal{R}^\nu_{pn}\right)~.\\
\label{Mmn1}
\end{equation}
Solving the above expression gives the first-order correction term as, 
\begin{equation}
   M^{(1)}_{mn} = \frac{e}{{\hbar}}\bigg[\tilde{E}_\mu g^{\omega_1}_{mn}\mathcal{R}^{\mu}_{mn} + \frac{\tilde{E}_{\mu \nu}}{2}g^{\omega_1}_{mn}\bigg(\sum_p \mathcal{R}^\mu_{mp}\mathcal{R}^\nu_{pn} \bigg)\bigg]~. 
\end{equation}
Here, we have defined $g^{\omega_1}_{mn} = [(\omega_1 - \omega_{mn})+i/\tau]^{-1}$, with $1/\tau = \xi $. The divergence appearing in the semiclassical theory is properly regularized by introducing a finite relaxation time $\tau$.

\section{Current Calculation}

\subsection{Distribution function}\label{App1}

In addition to the semiclassical EOMs, the nonequilibrium distribution function is needed to calculate the current. Within the relaxation time approximation \cite{ashcroft_book1976_solid}, the evolution of the distribution function is governed by the Boltzmann transport equation, which takes the form \cite{harris2004introduction},
\begin{equation}
\partial_t \bar{f}_n - \frac{e}{\hbar}\tilde{E}'_\mu\partial_\mu \bar{f}_n = - \frac{(\bar{f}_n-f_n)}{\tau}~.
\label{boltzapp}
\end{equation}
Here, $\bar{f}_n$ denotes the nonequilibrium distribution function, $f_n$ is the equilibrium Fermi-Dirac distribution, and $\tau$ is the relaxation time. The equilibrium distribution function is given by, $f_n = \left[ e^{(\epsilon_n-\mu)/k_BT}+1\right]^{-1}$ with $\mu$ being the chemical potential, $k_B$ the Boltzmann constant,
and $T$ being the absolute temperature. To solve Eq. \eqref{boltzapp} perturbatively, we expand the nonequilibrium distribution function in powers of the electric field, 
\begin{equation}
\bar{f}_n = f_n + f_n^{(1)} + f_n^{(2)} + \cdots,
\end{equation}
where $f_n^{(i)} \propto |\bm{\tilde{E}'}|^{i}$ for all $i > 0$,  denotes the $i$th order correction. Substituting this expansion into Eq.~\eqref{boltzapp} and collecting terms linear in the electric field yields the following first-order equation,
\begin{align}
\partial_t f_n^{(1)} +  \frac{f_n^{(1)}}{\tau} &= \frac{e}{\hbar}\tilde{E}'_\mu \partial_\mu f_n~.
\label{OneOrder}
\end{align}
Solving this linear differential equation, we get the following expression for $f_n^{(1)}$, 
\begin{align}
f_n^{(1)} =
\frac{ie}{\hbar}\left(\frac{\tilde{E}_\mu + \tilde{E}_{\mu \nu}r^\nu  }{\omega_1+i/\tau} \right)\partial_\mu f_n ~.
\label{neq1app}
\end{align}

\subsection{Linear responses}\label{App2}
We now calculate the linear current response in a conducting sample under a spatially inhomogeneous, time-dependent electric field. The local current density at position $\bm{R}$ is defined as \cite{Lapa2019}, 
\be
j^\mu (\bm{R}) = -e\sum_n\int_{\bm{r}}\int_{\bm{k}}\dot{r}_n^\mu \bar{f}_n\delta^D(\bm{r} - \bm{R})~. 
\ee
The band velocity combined with the first-order distribution function gives the linear Drude current as \cite{Smith2001},
\be 
 j^{\mu}_{ \rm{Dr}} (\bm{R})= -\frac{ie^2}{\hbar^2}\sum_{n} \int_{\bm{r}}\int_{\bm{k}}  \bigg[\dfrac{\tilde{E}_\nu + \tilde{E}_{\nu \lambda}r^\lambda  }{\omega_1+i/\tau} \bigg]\partial_\mu\epsilon_n\partial_\nu f_n\delta^D(\bm{r} - \bm{R})~,
\ee
where $\partial_\mu \equiv \partial / \partial k_\mu$, and $\tilde{E}_{\nu \lambda}$ accounts for the spatial variation of the electric field.
The current originating from the dipole of quantum metric and inhomogeneous electric field is given by,
\be
j^{\mu}_{\rm{QM}} (\bm{R})= -\frac{e^2}{2\hbar}\sum_{n,m} \int_{\bm{r}}\int_{\bm{k}}\tilde{E}_{\lambda \nu}{\partial_\mu \mathcal{G}^{\nu\lambda }_{mn}} f_n\delta^D(\bm{r} - \bm{R})~.
\ee 
The Berry curvature-induced anomalous velocity gives rise to an additional current \cite{Chang2023, Culcer_2024},
\be 
  j^{\mu}_{\rm{An}}(\bm{R}) = \frac{e^2}{\hbar}\sum_{n,m} \int_{\bm{r}}\int_{\bm{k}}   (\tilde{E}_\nu + \tilde{E}_{\nu \lambda} r^\lambda){\Omega}^{\mu\nu}_{nm}f_n \delta^D(\bm{r} - \bm{R})~.
\ee
\begin{figure*}[t!] 
    \includegraphics[width=\linewidth]{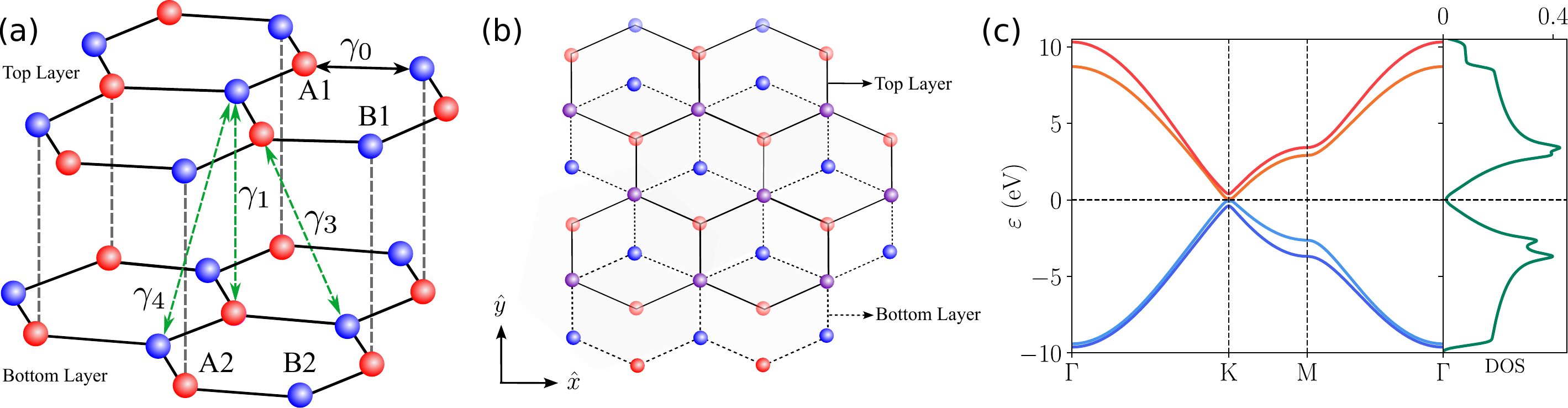}
    \caption{(a) The lattice structure of the Bernal stacked bilayer graphene showing all the hopping between the carbon atoms. A and B atoms are two sublattices. (b) The top view of the Bernal bilayer graphene. (c) The electronic band dispersion of the tight-binding bilayer graphene Hamiltonian along the high-symmetry path $\Gamma$–K–M–$\Gamma$ in the Brillouin zone. The corresponding density of states is shown alongside in units of $a^{-2}{\rm eV}^{-1}$. We have set the interlayer potential $\Delta=0.1$ eV.
\label{Fig3}}
\end{figure*}

In addition to these contributions, we have current originating from the time derivative of the first-order field-induced positional shift of the wavepacket, ($\partial_ta_n^\mu$). We decompose $a_n^\mu$ into homogeneous and inhomogeneous parts as,
\begin{equation}
a_n^\mu = a^{\mu,\rm{hom}}_n + a^{\mu,\rm{inhom}}_n~.
\end{equation}
The time derivative of the homogeneous position shift takes the form,
\bea
    \partial_t a^{\mu,\rm{hom}}_n &=& \frac{e}{\hbar}\partial_t\left[\tilde{E}_\nu \sum_{m} \frac{\mathcal{R}^\mu_{nm} \mathcal{R}^{\nu}_{mn} }{(\omega_1 - \omega_{mn}) +i /\tau}\right]~,\nn\\
    &=&
  \frac{e}{\hbar}\sum_{m}\left[\frac{\mathcal{R}^\mu_{nm} \mathcal{R}^{\nu}_{mn}\partial_t \tilde{E}_\nu  }{(\omega_1 - \omega_{mn}) +i \xi}+\frac{\mathcal{R}^\mu_{mn} \mathcal{R}^{\nu}_{nm}\partial_t \tilde{E}^*_\nu  }{(\omega_1 - \omega_{mn}) -i/\tau}\right]~,  
\eea
and the corresponding current is given by,
\bea
j_{\rm QGT}^\mu &=&-\frac{e^2}{\hbar}\sum_{n,m}\int_{\bm{r}}\int_{\bm{k}}\left[\frac{\mathcal{R}^\mu_{nm} \mathcal{R}^{\nu}_{mn}\partial_t \tilde{E}_\nu  }{(\omega_1 - \omega_{mn}) +i/\tau}+\frac{\mathcal{R}^\mu_{mn} \mathcal{R}^{\nu}_{nm}\partial_t \tilde{E}^*_\nu  }{(\omega_1 - \omega_{mn}) -i/\tau}\right]f_n\delta^D(\bm{r} - \bm{R})~,\nn\\
&=& 
-\frac{e^2}{\hbar}\sum_{n,m}\int_{\bm{r}}\int_{\bm{k}}\left[\frac{\mathcal{R}^\mu_{nm} \mathcal{R}^{\nu}_{mn}\partial_t \tilde{E}_\mu f_n }{(\omega_1 - \omega_{mn}) +i/\tau}-\frac{\mathcal{R}^\mu_{nm} \mathcal{R}^{\nu}_{mn}\partial_t \tilde{E}^*_\nu f_m }{(-\omega_1 - \omega_{mn}) +i/\tau}\right]\delta^D(\bm{r} - \bm{R})~,\nn\\
&=&
\frac{ie^2}{\hbar}\sum_{n,m}\int_{\bm{r}}\int_{\bm{k}}\left[\frac{\omega_1\mathcal{R}^\mu_{nm} \mathcal{R}^{\nu}_{mn}  f_{nm} }{(\omega_1 - \omega_{mn}) +i/\tau}\right]\tilde{E}_\nu\delta^D(\bm{r} - \bm{R})
~.  
\eea 
Here, the second line is obtained by interchanging $n\leftrightarrow m$, and the final form uses the identity 
$\sum_{\omega_1=\pm\omega}h(\omega_1)=\sum_{\omega_1=\pm\omega}h(-\omega_1)$. The time derivative of the inhomogeneous shift is given by,
\bea
    \partial_t a^{\mu,\rm{inhom}}_{n} &=& \frac{e}{\hbar} \partial_t \left[{\rm Re} \sum_{m,p}\tilde{E}_{\nu\lambda} \frac{\mathcal{R}^\mu_{nm}\mathcal{R}^\nu_{mp}\mathcal{R}^\lambda_{pn}}{(\omega_1 - \omega_{mn}) +i/\tau}\right]~,\nn\\
    &=&
    \frac{e}{2\hbar}  \sum_{m,p}\left[\frac{\mathcal{R}^\mu_{nm}\mathcal{R}^\nu_{mp}\mathcal{R}^\lambda_{pn}\partial_t \tilde{E}_{\nu\lambda} }{(\omega_1 - \omega_{mn}) +i/\tau} +\frac{\mathcal{R}^\mu_{mn}\mathcal{R}^\nu_{pm}\mathcal{R}^\lambda_{np}\partial_t \tilde{E}^*_{\nu\lambda} }{(\omega_1 - \omega_{mn}) -i/\tau} \right]~.
\eea
The correspondingly emerging higher-order connection (HOC) current for this is given by, 
\begin{eqnarray}
    j^\mu_{\rm{HOC}} 
    &=& -\frac{e^2}{2\hbar} \sum_{n,m,p}\int_{\bm{r}}\int_{\bm{k}} f_n \bigg[ \frac{\mathcal{R}^\mu_{nm}\mathcal{R}^\nu_{mp}\mathcal{R}^\lambda_{pn} (\partial_t\tilde{E}_{\nu \lambda})}{(\omega_1 - \omega_{mn}) +i/\tau} +\frac{\mathcal{R}^\mu_{mn}\mathcal{R}^\nu_{pm}\mathcal{R}^\lambda_{np} (\partial_t\tilde{E}^*_{\nu \lambda})}{(\omega_1 - \omega_{mn}) -i/\tau}\bigg]\delta^D(\bm{r} - \bm{R})~,\nn\\
    &=& -\frac{e^2}{2\hbar} \sum_{n,m,p} \int_{\bm{r}}\int_{\bm{k}} \bigg[ \frac{\mathcal{R}^\mu_{nm}\mathcal{R}^\nu_{mp}\mathcal{R}^{\lambda}_{pn} (\partial_t\tilde{E}_{\nu \lambda})}{(\omega_1 - \omega_{mn}) +i/\tau}f_n - \frac{\mathcal{R}^\mu_{nm}\mathcal{R}^\nu_{pn}\mathcal{R}^\lambda_{mp} (\partial_t\tilde{E}^*_{\nu \lambda})}{(-\omega_1 - \omega_{mn}) + i/\tau}f_m\bigg]\delta^D(\bm{r} - \bm{R}) ~,\nn\\
    &=&    \frac{ie^2}{2\hbar} \sum_{n,m,p}\int_{\bm{r}} \int_{\bm{k}} \frac{\omega_1  \tilde{E}_{\nu \lambda}}{(\omega_1 - \omega_{mn}) + i/\tau}
\bigg[ \mathcal{R}^\mu_{nm} \bigg(\mathcal{R}^\nu_{mp}\mathcal{R}^\lambda_{pn} f_n - \mathcal{R}^\lambda_{mp}\mathcal{R}^\nu_{pn} f_m \bigg)\bigg]\delta^D(\bm{r} - \bm{R})~.
\end{eqnarray}
The second line is obtained via the interchange $n\leftrightarrow m$, and we get the final form using the identity $\sum_{\omega_1=\pm\omega}h(\omega_1)=\sum_{\omega_1=\pm\omega}h(-\omega_1)$.
\end{widetext}

\section{Bernal stacked bilayer graphene}\label{App_C} 
To demonstrate HOC-driven responses (discussed in Sec.~\ref{section_three}) in a realistic system with tunable properties, we use Bernal-stacked bilayer graphene (BLG). BLG is a widely studied material where inversion symmetry can be easily broken by applying a vertical electric field. Its four-band low-energy model allows the HOC ( involving three bands) to be finite, and this enables BLG to support HOC-driven response. 

BLG has two single layers of graphene superimposed on top of each other.  As depicted in Fig. \ref{Fig3}  (a), the B2 atoms of the top layer lie directly above the A1 atoms of the bottom layer, forming the dimer sites, where the electronic orbitals are coupled together by interlayer coupling \cite{McCann_2013}. In contrast, the A2 and B1 atoms are situated above the centers of the hexagons in the opposite layers and remain electronically decoupled at low energies. The interlayer spacing in this configuration is $c \approx 3.35$ \AA.  

The tight binding Hamiltonian describing this system in the basis of (A1, B1, A2, B2) orbitals is \cite{Ghorai2025},
\begin{equation} \label{blg_ham}
    \mathcal{H} = \hbar v_F\begin{pmatrix}
        \Delta & -\gamma_0g_k & \gamma_4 g_k & -\gamma_3 g^*_k \\
        -\gamma_0 g^*_k & \Delta & \gamma_1 & \gamma_4 g_k \\
        \gamma_4 g^*_k & \gamma_1 &\Delta & -\gamma_0g_k \\
        -\gamma_3g_k & \gamma_4 g^*_k & -\gamma_0 g^*_k & -\Delta 
    \end{pmatrix}~.
\end{equation}
Here, $\Delta$ is the interlayer potential difference, and $v_F$ is the Fermi velocity. The intralayer nearest-neighbor hopping amplitude is $\gamma_0 = 3.16$ eV, while the interlayer couplings are $\gamma_1 = 0.381$ eV, $\gamma_3 = 0.38$ eV, and $\gamma_4 = 0.14$ eV. In Eq.~\eqref{blg_ham}, the diagonal $2 \times 2$ blocks represent the monolayer Hamiltonian, while the off-diagonal blocks account for the interlayer coupling. The structure factor for the in-plane non-dimeric hoppings is $g_k = \sum_{j=1}^3 e^{i \bm{k} \cdot \bm{\delta}_j}$, with the nearest neighbor vectors  $\bm{\delta}_1 = \left(0, a/\sqrt{3}\right)$, $\bm{\delta}_2 = \left(a/2, -a/(2\sqrt{3})\right)$, and $\bm{\delta}_3 = (-a/2, -a/(2\sqrt{3}))$. These vectors define the positions of the three nearest B atoms relative to an A atom, with the in-plane lattice constant being $a = 2.46$ \AA. 

\bibliography{ref.bib}
\end{document}